\documentclass[12pt]{article}
\usepackage{amssymb,amsmath,epsfig}
\allowdisplaybreaks
\begin{document}
\title{\bf Propagation of Polar Gravitational Waves in $f(R,T)$ Scenario}
\author{M. Sharif \thanks{msharif.math@pu.edu.pk} and Aisha Siddiqa
\thanks{aisha.siddiqa17@yahoo.com}\\
Department of Mathematics, University of the Punjab,\\
Quaid-e-Azam Campus, Lahore-54590, Pakistan.}

\date{}

\maketitle
\begin{abstract}
This paper investigates the propagation of polar gravitational waves
in the spatially flat FRW universe consisting of a perfect fluid in
the scenario of $R+2\lambda T$ model of $f(R,T)$ gravity ($\lambda$
being the model parameter). The spatially flat universe model is
perturbed via Regge-Wheeler perturbations inducing polar
gravitational waves and the field equations are formulated for both
unperturbed as well as perturbed spacetimes. We solve these field
equations simultaneously for the perturbation parameters introduced
in the metric, matter, and velocity in the radiation, as well as
dark energy, dominated phases. It is found that the polar
gravitational waves can produce changes in the background matter
distribution as well as velocity components in the radiation era
similar to general relativity case. Moreover, we have discussed the
impact of model parameter on the amplitude of gravitational waves.
\end{abstract}

\section{Introduction}

From the last few decades, researchers have been trying to disclose
the mystery behind accelerated expansion of the universe. On
theoretical grounds, they introduced various proposals to discuss
this issue including modification in the Einstein-Hilbert action.
Any change in the geometric part of this action leads to modified
theories of gravity, for example, $f(R)$ \cite{1a}, $f(G)$ \cite{1b}
and $f(R,T)$ \cite{2b}, where $R$ denotes the Ricci scalar, $G$
represents Gauss-Bonnet invariant and $T$ corresponds to the trace
of the energy-momentum tensor. Similarly, change in the matter part
produces different modified matter models \cite{1d}-\cite{1g}. There
is also another class of alternative theories to general relativity
(GR) called the teleparallel equivalent to GR \cite{1z} in which the
curvature scalar is replaced by the torsion scalar.

A direct and straight forward generalization of GR is $f(R)$
gravity. Harko \textit{et al}. \cite{2b} introduced a
curvature-matter coupled theory named as $f(R,T)$ gravity where the
dependence on $T$ is included due to the considerations of exotic
fluids or quantum effects. The extended curvature-matter coupling
leads to the existence of an extra force which can be associated
with the effects of dark matter. Different issues like cosmic
evolution \cite{2c1}-\cite{2c5}, laws of thermodynamics
\cite{2b1}-\cite{2b4}, stability analysis of compact objects
\cite{2d1}-\cite{2d9} as well as wormhole geometry
\cite{2e1}-\cite{2e4} have been studied in this context.

After the detection of gravitational waves (GWs) by LIGO-VIRGO
collaboration \cite{4a}-\cite{4d}, the study of GWs have gained much
importance. The detection and analysis of GWs produced by different
events can give information about structure formation and kinematics
of the universe. The geometrical orientation of these waves can be
investigated by exploring their polarization modes. In GR, a GW has
two polarization modes while in modified theories of gravity a GW
can have extra modes than GR. For example, it has been shown that a
GW has two extra modes than GR in $f(R)$ \cite{6c,6b} and $f(R,T)$
theories \cite{6d} (because $f(R,T)$ is reduced to $f(R)$ gravity in
vacuum). However, in $f(R,T^{\phi})$ gravity ($\phi$ stands for
scalar field), the number of polarization modes depends upon the
expression of $T^{\phi}$ \cite{6d}. We have examined the axially
symmetric dust with dissipation for gravitational radiation (energy
carried by GW) in $f(R)$ gravity \cite{6a}. It is found that
spinning fluid can radiate gravitationally in comparison to GR.

The study of GWs has been carried out by introducing different types
of perturbations in various cosmological backgrounds. Regge and
Wheeler \cite{5a} proposed perturbations in the Schwarzschild metric
that produce axial and polar waves. They analyzed these fluctuations
deducing that the Schwarzschild solution is stable under such
non-spherical disturbances. Zerilli \cite{5b} examined gravitational
radiation emitted by a black hole when it swallows a star by using
Regge-Wheeler perturbations and also corrected the polar wave
equation. Malec and Wyl\c{e}\.{z}ek \cite{5c} assumed these
Regge-Wheeler wavelike perturbations to investigate the GW
propagation in FRW background. They examined the validity of Huygens
principle for such GWs and found that this principle holds in
radiation dominated era but does not hold in matter dominated
universe. The perturbations proposed by Stewart and Walker
\cite{5r1} are used to study the GWs in Kantowski-Sachs \cite{5r2}
and locally rotationally symmetric class-II \cite{5r3} cosmological
backgrounds.

Viaggiu \cite{5c3} studied Regge-Wheeler perturbations (both axial
and polar) for de Sitter universe with the help of Laplace
transformation. Kulczycki and Malec \cite{5d} discussed the Huygens
principle for both axial as well as polar modes of GWs in FRW
background and examined the cosmological rotation induced by axial
GWs in radiation dominated universe \cite{5e}. These perturbations
have also been investigated using gauge-invariant quantities
\cite{5e1}-\cite{5e3}. Clarkson \textit{et al}. \cite{5e3} discussed
these cosmological perturbations in the background of
Lemaitre-Tolman metric.

The study of GWs via perturbation scheme (like Regge-Wheeler gauge)
has not been carried out in the context of modified theories of
gravity. Recently, we have explored the propagation of axial type
GWs (produced by odd wavelike perturbations) for the model
$f(R,T)=R+2\lambda T$ \cite{mr1} and found that GWs can generate
cosmological rotation when the wave profile exhibits a discontinuity
at the wave front. In this paper, we study the polar GWs induced by
the even type of wavelike perturbations in the similar background.
The paper has following format. In the next section, we give some
basic equations for the background metric in $f(R,T)$ scenario.
Section \textbf{3} consists of perturbation scheme, the
corresponding field equations and the simultaneous solution of these
equations for radiation and dark energy dominated universe. The
results are discussed in the last section.

\section{Background Cosmology}

The Weyl tensor vanishes for FRW cosmic models which implies
conformal flatness. Similar to axial waves, we consider the
background metric as the conformally flat FRW universe such that the
distortions are explicitly associated with polar GWs. The
conformally flat FRW model is defined as
\begin{equation}\label{1}
ds^{2}=a^{2}(\eta)(-d\eta^{2}+dr^{2}+r^{2}d\theta^{2}+r^{2}\sin^{2}\theta
d\phi^{2}),
\end{equation}
where $(\eta,r,\theta,\phi)$ are conformal coordinates with $\eta$
as conformal time. These coordinates are related to
$(t,r,\theta,\phi)$ via some conformal transformation. The relation
of conformal time with ordinary time is given by
\begin{equation}\label{1a}
\eta=\int\frac{dt}{a}.
\end{equation}
Here the conformal Hubble parameter $H$ is associated with the
ordinary Hubble parameter $\mathcal{H}$ by the relation
\begin{equation}\label{1b}
\mathcal{H}=\frac{H}{a}.
\end{equation}
The background matter has the following energy-momentum tensor
\begin{eqnarray}\label{3}
T_{\mu\nu}=(\rho_{0}+p_{0})V_{\mu}V_{\nu}+p_{0}g_{\mu\nu},
\end{eqnarray}
where $V_{\mu}$, $\rho_{0}$ and $p_{0}$ correspond to four velocity,
density and pressure, respectively.

The $f(R,T)$ gravity action
\begin{equation}\label{4}
S=\int
d^{4}x\sqrt{-g}\left[\frac{1}{16\pi}f(R,T)+\mathcal{L}_{m}\right],
\end{equation}
yields the following field equations
\begin{equation}\label{5}
f_{R}R_{\mu\nu}-\frac{1}{2}g_{\mu\nu}f-(\nabla_{\mu}\nabla_{\nu}-g_{\mu\nu}\Box)f_{R}=8\pi
T_{\mu\nu} -f_{T}(\Theta_{\mu\nu}+T_{\mu\nu}),
\end{equation}
where $\mathcal{L}_{m}$ is the matter Lagrangian density,
$f_{R}=\frac{\partial f}{\partial R},~f_{T}=\frac{\partial
f}{\partial T}$ and
$\Theta_{\mu\nu}=-2T_{\mu\nu}+\mathcal{L}_{m}g_{\mu\nu}$. We
consider the $f(R,T)$ gravity model as $f(R,T)=R+2\lambda T$ (with
$\lambda$ as a coupling constant called model parameter) \cite{2b},
which represents the simplest minimal curvature-matter coupling.
This model can be associated with $\Lambda$CDM model by considering
a trace dependent cosmological constant and also by $\Lambda(T)$
gravity proposed by Poplawski \cite{r10}. The choices for matter
Lagrangian density $\mathcal{L}_{m}$ are $p_{0}$ or $-\rho_{0}$ and
it is shown that these two densities yield the same results for
minimal curvature-matter coupling if the considered matter is
perfect fluid \cite{r11}. Further, assuming $\mathcal{L}_{m}=p_{0}$
and replacing $f(R,T)=R+2\lambda T$, we have the following form of
the field equations
\begin{equation}\label{5a}
G_{\mu\nu}=(8\pi+2\lambda) T_{\mu\nu}-2\lambda pg_{\mu\nu}+\lambda
Tg_{\mu\nu},
\end{equation}
leading to the following two independent field equations for the
metric (\ref{1})
\begin{eqnarray}\label{6}
3H^{2}=(8\pi+3\lambda)\rho_{0}a^{2}-\lambda p_{0}a^{2},\\\label{7}
-2\dot{H}-H^{2}=(8\pi+3\lambda)p_{0}a^{2}-\lambda \rho_{0}a^{2},
\end{eqnarray}
where an over dot indicates differentiation with respect to
conformal time $\eta$.

We evaluate the modified expressions of Hubble parameter, scale
factor and background density. For this reason, we formulate the
following differential equation in $H$ using the field equations
\begin{eqnarray}\label{n1}
-\dot{H}+H^{2}=(4\pi+\lambda)a^{2}(\rho_{0}+p_{0}).
\end{eqnarray}
To solve this equation for $H$, we can consider equation of state
(EoS) for different evolutionary eras of the universe. Using the EoS
for radiation era, $p_{0}=\frac{\rho_{0}}{3}$ as well as (\ref{6}),
Eq.(\ref{n1}) becomes
\begin{eqnarray}\nonumber
2\dot{H}+\frac{6\pi+\lambda}{3\pi+\lambda}H^{2}=0.
\end{eqnarray}
Its solution gives the following values of Hubble parameter and
scale factor
\begin{eqnarray}\label{24a}
H(\eta)=\frac{6\pi+2\lambda}{6\pi+\lambda}\frac{1}{\eta},\\\label{24aa}
a(\eta)=c_{1}\eta^{\frac{6\pi+2\lambda}{6\pi+\lambda}},
\end{eqnarray}
with $c_{1}$ is an integration constant. The covariant derivative of
the energy-momentum tensor is
\begin{eqnarray}\label{24c}
\nabla^{\mu}T_{\mu\nu}=\frac{f_{T}}{8\pi-f_{T}}\left[(T_{\mu\nu}+
\Theta_{\mu\nu})\nabla^{\mu}\ln
f_{T}+\nabla^{\mu}\Theta_{\mu\nu}-\frac{1}{2}g_{\mu\nu}\nabla^{\mu}T\right].
\end{eqnarray}
The first term within square brackets in the above equation vanishes
because $\nabla^{\mu}\ln f_{T}=0$ for our model and the last term
vanishes as $T=0$ for the radiation-dominated phase. Simplifying the
remaining terms, we obtain the following differential equation in
$\rho_{0}$ for radiation era
\begin{eqnarray}\nonumber
\dot{\rho}_{0}+3\frac{4\pi+\lambda}{3\pi+\lambda}H\rho_{0}=0,
\end{eqnarray}
whose solution is
\begin{eqnarray}\label{24b}
\rho_{0}=c_{2}a^{\frac{-3(4\pi+\lambda)}{3\pi+\lambda}},
\end{eqnarray}
$c_{2}$ is another integration constant. These expressions are
useful in onward calculations.

\section{Polar Perturbations and their Effects}

In this section, we study the polar form of wavelike perturbations
in an initially flat universe by considering Regge-Wheeler approach.
If $g^{(flat)}_{\mu\nu}$ denotes the background metric (i.e., FRW
universe) and the corresponding perturbations are denoted by
$h_{\mu\nu}$, then
\begin{equation}\label{1b}
g^{(perturb)}_{\mu\nu}=g^{(flat)}_{\mu\nu}+\hat{e}h_{\mu\nu}+O(\hat{e}^{2}),
\end{equation}
where $\hat{e}$ is a small parameter that measures the strength of
oscillations and the terms involving higher powers of $\hat{e}$ are
ignored. The polar or even perturbation matrix is \cite{5a}
\begin{equation}
h_{\mu\nu}=\left(
\begin{array}{cccc}
(\chi+\varphi)Y & \sigma Y & 0 & 0 \\
\sigma Y & (\chi+\varphi)Y  & 0 & 0 \\
0 & 0 & r^{2}\varphi Y & 0 \\
0 & 0 & 0 &  r^{2}\varphi Y\sin^{2}\theta \\
\end{array}
\right),
\end{equation}
where $\chi$, $\varphi$, $\sigma$ are functions of $\eta$ and $r$
while $Y(\theta)=Y_{lm}(\theta)$ (where $l$ denotes the angular
momentum and $m$ corresponds to its projection on $z$-axis) are the
spherical harmonics. Here we take $m=0$ \cite{5a} as well as
$l=2,3,...$ for wavelike solution. Consequently, the perturbed
spacetime can be written as
\begin{eqnarray}\nonumber
ds^{2}&=&(-a^{2}+\hat{e}(\chi+\varphi)Y)d\eta^{2}+2\hat{e}\sigma Y
d\eta
dr+(a^{2}+\hat{e}(\chi+\varphi))dr^{2}\\\label{2}&+&r^{2}(a^{2}+\hat{e}\varphi
Y)(d\theta^{2}+\sin^{2}d\phi^{2})+O(\hat{e}^{2}).
\end{eqnarray}

The perturbations in matter quantities are taken as \cite{5d}
\begin{eqnarray}\label{3b}
\rho&=&\rho_{0}(1+\hat{e}\Delta(\eta,r)Y)+O(\hat{e}^{2}),\\\label{3b1}
p&=&p_{0}(1+\hat{e}\Pi(\eta,r)Y)+O(\hat{e}^{2}).
\end{eqnarray}
In the perturbed scenario, the fluid also suffers distortions and
may become non-comoving such that the perturbed four velocity has
the components \cite{5d}
\begin{eqnarray}\label{v1}
V_{0}&=&\frac{2g^{(0)}_{00}+\hat{e}k_{00}}{2a(\eta)}+O(\hat{e}^{2}),\\\label{v2}
V_{1}&=&\hat{e}a(\eta)w(\eta,r)Y+O(\hat{e}^{2}),\\\label{v3}
V_{2}&=&\hat{e}v(\eta,r)Y'+O(\hat{e}^{2}),\\\label{v4}
V_{3}&=&\hat{e}\sin\theta u(\eta,r)Y'+O(\hat{e}^{2}),
\end{eqnarray}
with $V_{\alpha}V^{\alpha}=-1+O(\hat{e}^{2})$, i.e., the above
defined form of velocity components satisfy the normalization
condition with arbitrary functions $w$, $v$ and $u$. We first
formulate the perturbed field equations and then discuss the
evolution of the unknown perturbation parameters in the
curvature-matter coupling background. The perturbed field equations
are obtained as
\begin{eqnarray}\nonumber
&&-2\varphi''-\frac{4}{r}\varphi'+6H\dot{\varphi}-6H^{2}\varphi+2\frac{l(l+1)}{r^{2}}\varphi
+\frac{2}{r}\chi'+2H\dot{\chi}+\frac{2\chi}{r^{2}}+2\chi
H^{2}\\\label{f1}&&-\frac{l(l+1)}{r^{2}}\chi
-4H\sigma'-\frac{8H}{r}\sigma
=2a^{4}[(8\pi+3\lambda)\rho_{0}\Delta-\lambda p_{0}\Pi],\\\nonumber
&&-2\ddot{\varphi}+2H\dot{\varphi}+2H^{2}\varphi-2H\dot{\chi}+\frac{2\chi'}{r}
+\frac{\chi}{r^{2}}[l(l+1)-2]
-4\dot{H}\chi+2H^{2}\chi\\\label{f2}&&=2a^{4}[(8\pi+3\lambda)p_{0}\Pi-\lambda
\rho_{0}\Delta],\\\label{f3} &&\chi'=\dot{\sigma},\\\nonumber
&&-\frac{\ddot{\chi}}{2}-\frac{\chi''}{2}-\frac{\chi'}{r}+\chi(H^{2}-\dot{H})
-\ddot{\varphi}+H\dot{\varphi}+H^{2}\varphi+\dot{\sigma}'+\frac{\dot{\sigma}}{r}
\\\label{f4}&&=a^{4}[
(8\pi+3\lambda)p_{0}\Pi-\lambda
\rho_{0}\Delta],\\\nonumber
&&2\dot{\varphi}'+2H\varphi'+2H\chi'-\frac{2}{r}\dot{\chi}+\frac{4H\chi}{r}-2H^{2}\sigma
-\frac{l(l+1)}{r}\sigma\\\label{f5}&&=a^{2}w(8\pi+2\lambda)(\rho_{0}+p_{0}),\\\label{f6}
&&2\dot{\varphi}-2H\varphi+\dot{\chi}-\sigma'=a^{3}v(8\pi+2\lambda)(\rho_{0}+p_{0}),
\end{eqnarray}
where prime shows differentiation with respect to $r$. During
simplification of the above equations, we have used the unperturbed
field equations as well as the following relation \cite{5d}
\begin{equation}\nonumber
\partial_{\theta}\partial_{\theta}Y=-l(l+1)Y-\cot\theta
\partial_{\theta}Y.
\end{equation}
Equations (\ref{f1})-(\ref{f4}) relate geometrical and material
perturbations while (\ref{f5}) and (\ref{f6}) describe the
deformation in velocity components due to polar perturbations. Also,
we have $u=0$ because $G_{03}=0$. Using Eqs.(\ref{f2}) and
(\ref{f3}) in (\ref{f4}), we obtain
\begin{eqnarray}\label{f7}
\ddot{\chi}-\chi''-2H\dot{\chi}+2\frac{\chi'}{r}-2\dot{H}\chi+\frac{\chi}{r^{2}}[l(l+1)-2]=0,
\end{eqnarray}
which looks similar to Eq.(33) in \cite{5d} but here the values of
$H$ and $\dot{H}$ depend on the model parameter $\lambda$.

To solve the above equation, we assume that
\begin{eqnarray}\label{s1}
\chi(\eta,r)=ra(\eta)q(\eta,r),
\end{eqnarray}
which gives
\begin{eqnarray}\label{s2}
\ddot{q}-q''+\left[\frac{l(l+1)}{r^{2}}-\dot{H}-H^{2}\right]q=0.
\end{eqnarray}
Replacing $l=2$ (for wavelike solution) and using Eq.(\ref{24a}), we
have
\begin{eqnarray}\label{s2}
\ddot{q}-q''+\left[\frac{6}{r^{2}}+\frac{A_{1}}{\eta^{2}}\right]q=0,
\end{eqnarray}
where
$A_{1}=\frac{6\pi+2\lambda}{6\pi+\lambda}-\left(\frac{6\pi+2\lambda}{6\pi+\lambda}\right)^{2}$.
This is the wave equation and its solution can be obtained through
separation of variables technique. For this purpose, we assume
$q(\eta,r)=\mathbb{T}(\eta)\mathbb{R}(r)$ and the separation
constant $-A_{2}^{2}$, it follows that
\begin{eqnarray}\label{27}
\mathbb{R}''+\left(A_{2}^{2}-\frac{6}{r^{2}}\right)\mathbb{R}&=&0,\\\label{28}\ddot{\mathbb{T}}
+\left(A_{2}^{2}+\frac{A_{1}}{\eta^{2}} \right)\mathbb{T}&=&0.
\end{eqnarray}
The solutions of these differential equations are given by,
respectively.
\begin{eqnarray}\nonumber
\mathbb{R}(r)&=&\sqrt{\frac{2}{A_{2}\pi}}c_{3}\left(\frac{-3\cos
A_{2} r}{A_{2} r}-\sin A_{2} r+\frac{3\sin A_{2}
r}{A_{2}^{2}r^{2}}\right)\\\label{27a}&+&\sqrt{\frac{2}{A_{2}\pi}}c_{4}\left(\frac{-3\cos
A_{2} r}{A_{2}^{2}r^{2}}-\frac{3\sin A_{2} r}{A_{2} r}+\cos A_{2}
r\right),\\\nonumber
\mathbb{T}(\eta)&=&c_{5}\sqrt{\eta}J_{\frac{\sqrt{1-4A_{1}}}{2}}(A_{2}\eta)
+c_{6}\sqrt{\eta}Y_{\frac{\sqrt{1-4A_{1}}}{2}}(A_{2}\eta),
\end{eqnarray}
where $c_{i}'s$ for $i=3,4,5,6$ are constants of integration,
$J_{\frac{\sqrt{1-4A_{1}}}{2}}(A_{2}\eta)$ and
$Y_{\frac{\sqrt{1-4A_{1}}}{2}}(A_{2}\eta)$ are Bessel functions of
first and second kind, respectively. Their expressions are
\begin{eqnarray}\nonumber
J_{\frac{\sqrt{1-4A_{1}}}{2}}(A_{2}\eta)&=&
\sum_{m=0}^{\infty}\frac{(-1)^{m}}{m!\Gamma\left(m+\alpha+1\right)}
\left(\frac{\eta}{2}\right)^{2m+\alpha},\\\nonumber
Y_{\frac{\sqrt{1-4A_{1}}}{2}}(A_{2}\eta)&=&
\frac{J_{\alpha}(A_{2}\eta)\cos\alpha\pi-J_{-\alpha}(A_{2}\eta)}{\sin
A_{2}\eta},
\end{eqnarray}
with $\alpha=\frac{\sqrt{1-4A_{1}}}{2}$.

Inserting $\mathbb{T}(\eta)$ and $\mathbb{R}(r)$ in
$q(\eta,r)=\mathbb{T}(\eta)\mathbb{R}(r)$, we obtain $q$ which leads
Eq.(\ref{s1}) to the expression of $\chi$ as
\begin{eqnarray}\nonumber
\chi(\eta,r)&=&\frac{-c_{1}\sqrt{\frac{2}{\pi}}}{rA_{2}^{\frac{5}{2}}}
\eta^{\frac{18\pi+5\lambda}{12\pi+2\lambda}}
\{(3A_{2} rc_{3}+3c_{4}-A_{2}^{2}r^{2}c_{4})\cos A_{2} r+(3A_{2}
rc_{4}-3c_{3}\\\nonumber&+&A_{2}^{2}r^{2}c_{3})\sin A_{2} r\}
\left\{c_{5}\sqrt{\eta}J_{\frac{\sqrt{1-4A_{1}}}{2}}(A_{2}\eta)
+c_{6}\sqrt{\eta}Y_{\frac{\sqrt{1-4A_{1}}}{2}}(A_{2}\eta)\right\},\\\label{s11}
\end{eqnarray}
Moreover, Eq.(\ref{f3}) yields
\begin{eqnarray}\label{29}
\sigma(\eta,r)=\mathcal{B}(r)+\int_{\eta_{0}}^{\eta}
\chi'(\tau,r)d\tau,
\end{eqnarray}
where $\eta_{0}$ corresponds to conformal time at the hypersurface
generating GWs. Assuming $\sigma(\eta_{0},r)=0$, we have
$\mathcal{B}(r)=0$ and using Eq.(\ref{s1}), $\sigma$ becomes
\begin{eqnarray}\nonumber
\sigma(\eta,r)=(r\mathbb{R}(r))'\int_{\eta_{0}}^{\eta}
a(\tau)\mathbb{T}(\tau)d\tau,
\end{eqnarray}
which gives
\begin{eqnarray}\nonumber
\sigma(\eta,r)&=&\frac{-c_{1}\sqrt{r}2^{\frac{-1-\sqrt{1-4A_{1}}}{2}}}{\sqrt{\pi}(A_{2}
r)^{\frac{5}{2}}}\eta^{\frac{20\pi+7\lambda}{12\pi+2\lambda}}[(-3A_{2}
rc_{3}+A_{2}^{3}r^{3}c_{3}-3c_{4}+2A_{2}^{2}r^{2}c^{4})\\\nonumber&\times&\cos
A_{2} r+(3c_{3}-2A_{2}^{2}r^{2}c_{3}-3A_{2}
rc_{4}+A_{2}^{3}r^{3}c_{4})\sin A_{2} r]\\\nonumber&\times&
\left[(A_{2}\eta)^{\sqrt{1-4A_{1}}}\left(c_{5}+c_{6}\cot\left(\frac{\sqrt{1-4A_{1}}\pi}{2}\right)\right)
\right.\\\nonumber&\times&\left.\Gamma\left[\frac{6(5+\sqrt{1-4A_{1}})\pi+(7+\sqrt{1-4A_{1}})\lambda}{4(6\pi+\lambda)}\right]
\right.\\\nonumber&\times&\left.\text{HypergeometricPFQRegularized}
\left[\frac{6(5+\sqrt{1-4A_{1}})\pi}{4(6\pi+\lambda)}
\right.\right.\\\nonumber&+&\left.\left.\frac{(7+\sqrt{1-4A_{1}})\lambda}{4(6\pi+\lambda)},
\frac{1}{2}(2+\sqrt{1-4A_{1}}),
\frac{6(9+\sqrt{1-4A_{1}})\pi}{4(6\pi+\lambda)}
\right.\right.\\\nonumber&+&\left.\left.\frac{(11+\sqrt{1-4A_{1}})\lambda}{4(6\pi+\lambda)},
\frac{-A_{2}^{2}\eta^{2}}{4}\right]-
2^{\sqrt{1-4A_{1}}}c_{6}\csc\left(\frac{\sqrt{1-4A_{1}}\pi}{2}\right)\right.\\\nonumber&\times&
\left.\Gamma\left[\frac{(2-\sqrt{1-4A_{1}})}{4}+\frac{(18\pi+5\lambda)}{4(6\pi+\lambda)}\right]
\right.\\\nonumber&\times&\left.\text{HypergeometricPFQRegularized}
\left[\frac{(2-\sqrt{1-4A_{1}})}{4}\right.\right.\\\nonumber&+&\left.\left.
\frac{(18\pi+5\lambda)}{4(6\pi+\lambda)},1-\frac{\sqrt{1-4A_{1}}}{2},
\frac{(2-\sqrt{1-4A_{1}})}{4}\right.\right.\\\label{29a}&+&\left.\left.\frac{(18\pi+5\lambda)}{4(6\pi+\lambda)}
,\frac{-A_{2}^{2}\eta^{2}}{4}\right]\right].
\end{eqnarray}
Here HypergeometricPFQregularized is the regularized generalized
hypergeometric function.
\begin{figure}
\center\epsfig{file=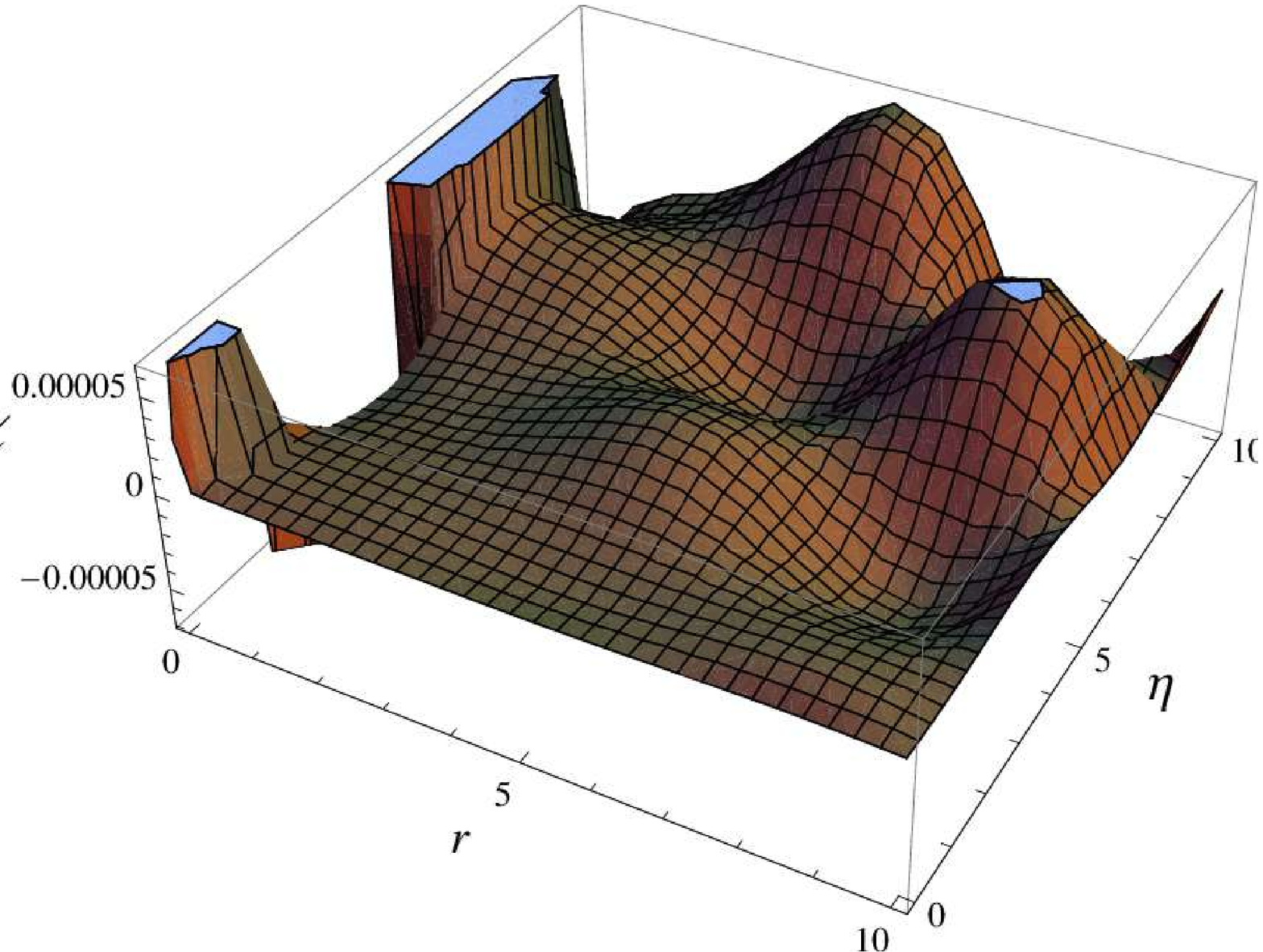,width=0.55\linewidth}\center\epsfig{file=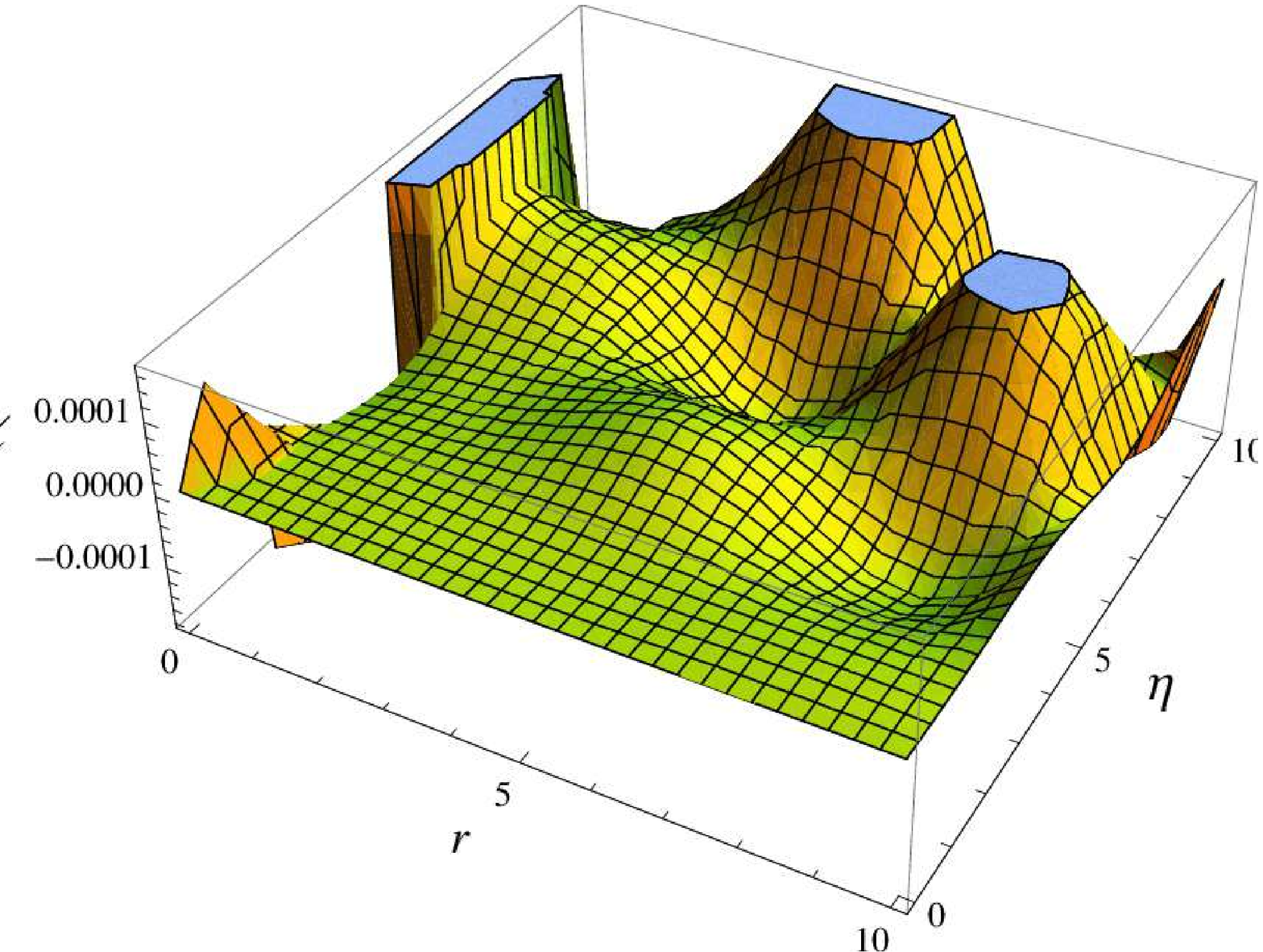,width=0.55\linewidth}
\center\epsfig{file=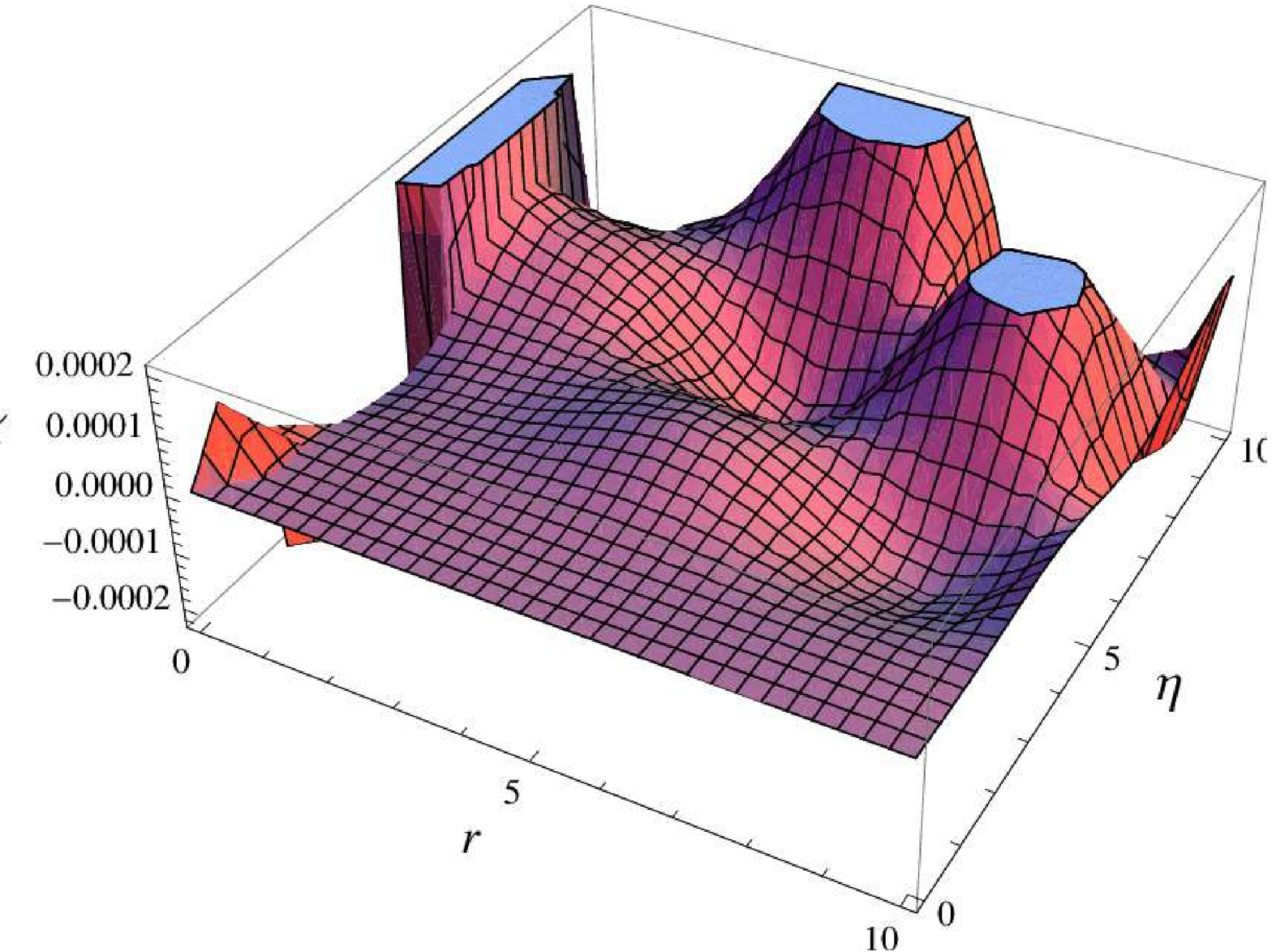,width=0.55\linewidth} \caption{Plots of
$\chi$ versus $r$ and $\eta$ for $A_{1}=-1$, $c_{3}=0.01$,
$c_{4}=0.01$, $c_{5}=0.01$, $c_{6}=0.01$, $\lambda=0$ (brown),
$\lambda=3(1+\sqrt{5})\pi$ (yellow) and $\lambda=3(1+\sqrt{5})\pi$
(pink).}
\end{figure}

To find the remaining metric perturbation coefficient $\varphi$, we
consider the speed of sound formula $v_{s}^{2}=\frac{\partial
p}{\partial\rho}$ which gives $\Delta\rho_{0}=\Pi p_{0}$. Using this
relation (between $\Delta$ and $\Pi$) and then comparing
Eqs.(\ref{f1}), (\ref{f2}), we have
\begin{eqnarray}\nonumber
&&-2A_{3}\varphi''+2A_{4}\ddot{\varphi}+(6A_{3}-2A_{4})\dot{\varphi}H-\frac{4A_{3}}{r}\varphi'+
\left\{A_{3}\left(\frac{12}{r^{2}}-6H^{2}\right)\right.\\\nonumber&&\left.-2H^{2}A_{4}\right\}\varphi+
\frac{2}{r}(A_{3}-A_{4})\chi'+2H(A_{3}+A_{4})\dot{\chi}+
\left\{A_{3}\left(-\frac{4}{r^{2}}+2H^{2}\right)\right.\\\label{29b}&&\left.-A_{4}
\left(\frac{2}{r}-\frac{4}{r^{2}}-4\dot{H}+2H^{2}\right)\right\}\chi-4HA_{3}\sigma'-\frac{8H}{r}A_{3}\sigma=0,
\end{eqnarray}
with $A_{3}=8\pi v_{s}^{2}+(3v_{s}^{2}-1)\lambda$ and $A_{4}=8\pi
+(3-v_{s}^{2})\lambda$. Replacing the values of $\chi$ and $\sigma$
in the above equation, we can obtain the evolution equation of
$\varphi$ which will be a hyperbolic differential equation and may
be solved using some suitable initial as well as boundary
conditions. The first two perturbed field equations describe the
evolution of matter fluctuations and the last two can depict the
disturbances in velocity components. All these evolution equations
for perturbation parameters contain the curvature-matter coupling
parameter $\lambda$ implying that their evolution will be affected
by $\lambda$. To observe the effects of coupling on the propagation
of GWs, we plot the metric perturbation parameter $\chi$ for three
different values of $\lambda$ as shown in Figure \textbf{1}. The
constant $A_{1}$ is related with $\lambda$ via the relation
$A_{1}=\frac{6\pi+2\lambda}{6\pi+\lambda}-\left(\frac{6\pi+2\lambda}{6\pi+\lambda}\right)^{2}$.
Thus we choose the value of $A_{1}$ such that the corresponding
value of $\lambda$ lies in the range $\lambda>-4\pi$ which comes
from the viability or stability criteria for any $f(R,T)$ gravity
model \cite{mr8}. For the assumed values of all constants as well as
ranges of independent parameters, the maximum amplitude of wave is
approximately $0.00005$ for $\lambda=0$ (GR case), $0.0001$ for
$3(1+\sqrt{5})\pi$ and $0.0002$ for $6(1+\sqrt{5})\pi$. It can
easily be noted that the increase in the value of $\lambda$ causes
an increase in the maximal amplitude of GW.

Now, we extend our discussion for dark energy dominated phase of
cosmic evolution. For this purpose, we substitute the EoS
$p_{0}=-\rho_{0}$ in the field equations and obtain the following
evolution equations
\begin{eqnarray}\nonumber
&&-2\varphi''-\frac{4}{r}\varphi'+6H\dot{\varphi}-6H^{2}\varphi+2\frac{l(l+1)}{r^{2}}\varphi
+\frac{2}{r}\chi'+2H\dot{\chi}+\frac{2\chi}{r^{2}}+2\chi
H^{2}\\\label{f11}&&-\frac{l(l+1)}{r^{2}}\chi
-4H\sigma'-\frac{8H}{r}\sigma
=2a^{4}[(8\pi+3\lambda)\Delta+\lambda\Pi] \rho_{0},\\\nonumber
&&-2\ddot{\varphi}+2H\dot{\varphi}+2H^{2}\varphi-2H\dot{\chi}+\frac{2\chi'}{r}
+\frac{\chi}{r^{2}}[l(l+1)-2]
-4\dot{H}\chi+2H^{2}\chi\\\label{f12}&&=-2a^{4}[(8\pi+3\lambda)\Pi+\lambda\Delta]
\rho_{0},
\\\label{f13}&&\chi'=\dot{\sigma},\\\nonumber
&&-\frac{\ddot{\chi}}{2}-\frac{\chi''}{2}-\frac{\chi'}{r}+\chi(H^{2}-\dot{H})
-\ddot{\varphi}+H\dot{\varphi}+H^{2}\varphi+\dot{\sigma}'+\frac{\dot{\sigma}}{r}
\\\label{f14}&&=-a^{4}[(8\pi+3\lambda)\Pi+\lambda
\Delta]\rho_{0},\\\label{f15}
&&2\dot{\varphi}'+2H\varphi'+2H\chi'-\frac{2}{r}\dot{\chi}+\frac{4H\chi}{r}-2H^{2}\sigma
-\frac{l(l+1)}{r}\sigma=0,\\\label{f16}
&&2\dot{\varphi}-2H\varphi+\dot{\chi}-\sigma'=0,
\end{eqnarray}
Here, the Hubble parameter and background density have the following
expressions
\begin{eqnarray}\nonumber
H(\eta)=\frac{1}{\hat{c}_{1}-\eta}, \quad
\rho_{0}=\frac{3}{4(2\pi+\lambda)\hat{c}_{2}},
\end{eqnarray}
with $\hat{c}_{1}$ and $\hat{c}_{2}$ as constants of integration.
Equation (\ref{f14}) yields the same evolution equation for $\chi$
as in the radiation era. Consequently, we have the same expressions
for the metric perturbation parameters $\chi$ and $\sigma$ (for
$\hat{c}_{1}=0$) with $A_{1}=-2$. We also find the expression of
$\varphi$ from Eq.(\ref{f16}) but that is too much lengthy to
present here. The expressions of $\Delta$ and $\Pi$ can be found
from Eqs.(\ref{f11}) and (\ref{f12}). However, the velocity
perturbation parameters do not appear in these field equations.

\section{Concluding Remarks}

Gravitational waves are the outcome of some cosmic events such as
big bang, gravitational collapse and mergence of two black holes or
ultra dense neutron stars. These waves flow over the Earth
continuously but our instruments are not yet too sensitive to detect
most of them individually. A very few of these signals that can be
assigned to a single event have been observed by the latest laser
interferometers. These collisions of compact objects also produce
stochastic background of GWs and advanced LIGO-Virgo interferometers
are sensitive to these signals with a certain frequency and wave
amplitude \cite{c1}. If the GW causes a displacement in the initial
positions (i.e., the positions before passing the GW) of freely
falling particles which remains permanently after the GW passed, we
speak of the GW memory effect. A GW without this effect can cause an
oscillatory deformation in a detector while with memory effect can
cause a permanent deformation in a true free falling detector
\cite{Tr3}.

Gravitational waves have also been studied extensively in modified
theories of gravity. It is expected that the Earth based network of
interferometers could provide constraints for these theories by
analyzing different properties of GWs. Cai \textit{et al.}
\cite{Tr6} and Li \textit{et al.} \cite{Tr7} discussed some
constraints for teleparallel theory using the effective field theory
approach. Gong et al. \cite{Tr8} worked out to constrain the scalar
tensor theory in view of recent GW observations.

This paper investigates the propagation of polar GWs in $f(R,T)$
theory of gravity. The polar GWs are assumed by perturbing the flat
FRW metric through polar (or even) waves in Regge-Wheeler gauge. The
corresponding matter and four velocity are also perturbed such that
the four velocity may be non-comoving. We find the perturbing
parameters via the field equations. The evolution equation of $\chi$
is a wave equation yielding its value (\ref{s11}) which in turn
gives $\sigma$ (\ref{29a}). Equation (\ref{29b}) leads to a
hyperbolic differential equation in $\varphi$ after replacing the
expressions of $\chi$ and $\sigma$. This can be solved using
suitable initial and boundary conditions as well as choosing
appropriate free parameters. The parameters $v$ and $w$ can be
explicitly obtained from Eqs.(\ref{f5}) and (\ref{f6}) while $u$
vanishes.

All the equations as well as obtained expressions have the terms
involving model parameter $\lambda$ indicating that the
curvature-matter coupling affect the evolution of perturbation
parameters and consequently the propagation of polar waves in
radiation dominated era. We have also investigated the propagation
of polar waves in dark energy dominated universe. It is observed
that the increase in $\lambda$ can enhance wave amplitude, i.e., we
can expect GWs with higher amplitudes if the value of
curvature-matter coupling constant increases. The analysis of
observed GW signals by LIGO-VIRGO collaboration shows that masses of
the merging black holes are directly related with energy radiated by
the coalesce as well as the amplitude of resulting GW. For
$f(R,T)=R+2\lambda T$ gravity model, it has also been observed that
if the value of $\lambda$ is increased, it increases the mass range
of the corresponding compact object \cite{2d4,2d8}. Moreover, when
these compact objects are in binaries and their mergence produces
GWs then the massive objects produce GWs of greater amplitude. Hence
our conclusion that amplitude increases with $\lambda$ is in
agreement with the literature.

In GR \cite{5d}, it is proved that vanishing of material
perturbations lead to vanishing of polar GWs, i.e., if polar waves
are passing through the spacetime the corresponding mater must also
be perturbed. In the absence of material perturbations, i.e., for
$\Delta=\Pi=v=w=0$ we obtain the same perturbed field equations as
in GR. The only difference is in the value of $H$ which has no
impact on the above mentioned fact. Thus, we can conclude that for
$f(R,T)=R+2\lambda T$ gravity model, there are no polar GWs when
there are no matter perturbations which is consistent with GR.

For axial waves, we have found the expressions of perturbation
parameters which show the effect of $f(R,T)=R+2\lambda T$ gravity on
the propagation of axial GW \cite{mr1}. To find these perturbations,
we use the constraint that matter perturbations vanish for axial
waves which are originally non-zero in $f(R,T)$ gravity in contrast
to GR. However, for polar waves, such an assumption leads to
vanishing of polar GWs. For polar waves, we have checked the effects
of $R+2\lambda T$ gravity model by plotting the metric perturbation
function $\chi$ for different values of $\lambda$.

\vspace{1.0cm}

{\bf Acknowledgment}

\vspace{0.25cm}

One of us (AS) would like to thank the Higher Education Commission,
Islamabad, Pakistan for its financial support through the {\it
Indigenous Ph.D. 5000 Fellowship Program Phase-II, Batch-III}.

\end{document}